\documentstyle[12pt,epsf]{article}

\catcode`\@=11
\long\def\@makefntext#1{
\protect\noindent \hbox to 3.2pt {\hskip-.9pt
$^{{\ninerm\@thefnmark}}$\hfil}#1\hfill}                

\def\@makefnmark{\hbox to 0pt{$^{\@thefnmark}$\hss}}  

\def\ps@myheadings{\let\@mkboth\@gobbletwo
\def\@oddhead{\hbox{}
\rightmark\hfil\ninerm\thepage}
\def\@oddfoot{}\def\@evenhead{\ninerm\thepage\hfil
\leftmark\hbox{}}\def\@evenfoot{}
\def\sectionmark##1{}\def\subsectionmark##1{}}

\renewcommand{\thefootnote}{\fnsymbol{footnote}}

\def\sectionc{\@startsection {section}{1}{\z@}{-3.5ex plus -1ex minus 
    -.2ex}{2.3ex plus .2ex}{\bf }}
\def\subsectionc{\@startsection{subsection}{2}{\z@}{-3.25ex plus -1ex minus 
   -.2ex}{1.5ex plus .2ex}{\it }}
\renewcommand{\section}[1]{\sectionc{#1}\hspace*{\parindent}}
\renewcommand{\subsection}[1]{\subsectionc{#1}\hspace*{\parindent}}
\newcounter{appendixc}
\newcounter{subappendixc}[appendixc]
\newcounter{subsubappendixc}[subappendixc]

\renewcommand{\appendix}[1] {\vspace*{0.6cm}
        \refstepcounter{appendixc}
        \setcounter{figure}{0}
        \setcounter{table}{0}
        \setcounter{equation}{0}
        \renewcommand{\thefigure}{\Alph{appendixc}.\arabic{figure}}
        \renewcommand{\thetable}{\Alph{appendixc}.\arabic{table}}
        \renewcommand{\theappendixc}{\Alph{appendixc}}
        \renewcommand{\theequation}{\Alph{appendixc}.\arabic{equation}}
        \noindent{\bf Appendix \theappendixc #1}\par\vspace*{0.4cm}}

\def\abstracts#1{{
        \centering{\begin{minipage}{13.2truecm}
        \footnotesize\baselineskip=13pt\noindent
        \parindent=0pt #1
        \end{minipage}}\par}}

\renewenvironment{thebibliography}[1]
        {\begin{list}{\arabic{enumi}.}
        {\usecounter{enumi}\setlength{\parsep}{0pt}
\setlength{\leftmargin 0.75cm}{\rightmargin 0pt}
         \setlength{\itemsep}{0pt} \settowidth
        {\labelwidth}{#1.}\sloppy}}{\end{list}}

\newcounter{itemlistc}
\newcounter{romanlistc}
\newcounter{alphlistc}
\newcounter{arabiclistc}

\newcommand{\fcaption}[1]{
        \refstepcounter{figure}
        \setbox\@tempboxa = \hbox{\footnotesize Figure~\thefigure. #1}
        \ifdim \wd\@tempboxa > 6in
           {\begin{center}
        \parbox{6in}{\footnotesize\baselineskip=13pt Figure~\thefigure. #1}
            \end{center}}
        \else
             {\begin{center}
             {\footnotesize Figure~\thefigure. #1}
              \end{center}}
        \fi}

\newcommand{\tcaption}[1]{
        \refstepcounter{table}
        \setbox\@tempboxa = \hbox{\footnotesize Table~\thetable. #1}
        \ifdim \wd\@tempboxa > 6in
           {\begin{center}
        \parbox{6in}{\footnotesize\baselineskip=13pt Table~\thetable. #1}
            \end{center}}
        \else
             {\begin{center}
             {\footnotesize Table~\thetable. #1}
              \end{center}}
        \fi}

\def\@citex[#1]#2{\if@filesw\immediate\write\@auxout
        {\string\citation{#2}}\fi
\def\@citea{}\@cite{\@for\@citeb:=#2\do
        {\@citea\def\@citea{,}\@ifundefined
        {b@\@citeb}{{\bf ?}\@warning
        {Citation `\@citeb' on page \thepage \space undefined}}
        {\csname b@\@citeb\endcsname}}}{#1}}

\newif\if@cghi
\def\cite{\@cghitrue\@ifnextchar [{\@tempswatrue
        \@citex}{\@tempswafalse\@citex[]}}
\def\citelow{\@cghifalse\@ifnextchar [{\@tempswatrue
        \@citex}{\@tempswafalse\@citex[]}}
\def\@cite#1#2{{$\null^{#1}$\if@tempswa\typeout
        {IJCGA warning: optional citation argument
        ignored: `#2'} \fi}}

 1
 1
 1

\font\ninerm=cmr9

\def\thefootnote{\fnsymbol{footnote}}
\def\bea {\begin{eqnarray}}
\def\eea {\end{eqnarray}}
\def\be {\begin{equation}}
\def\ee {\end{equation}}
\def\ben{\begin{enumerate}}
\def\een{\end{enumerate}}
\def\bi{\begin{itemize}}
\def\ei{\end{itemize}}

\def\F{{\cal F}}

\def\prl {{\it Phys. Rev. Lett.\ }}

\def\pr {{\it Phys. Rev.\ }}
\def\np {{\it Nucl. Phys.\ }}

\def\GA{G_{\mbox{\tiny A}}}
\def\GV{G_{\mbox{\tiny V}}}

\def\mA{m_{\mbox{\tiny A}}}

\def\mZ{m_{\mbox{\tiny Z}}}

\def\MF{M_{\mbox{\tiny F}}}
\def\MGT{M_{\mbox{\tiny GT}}}

\def\mids{\! \mid \! }

\begin{document}

\centerline{\normalsize\bf DEGREE TO WHICH CVC IS ESTABLISHED 
THROUGH}
\baselineskip=15pt
\centerline{\normalsize\bf BETA DECAY ALONE}

\vspace*{0.6cm}
\centerline{\footnotesize E. HAGBERG, J.C. HARDY, V.T. KOSLOWSKY, 
G. SAVARD and I.S. TOWNER}
\baselineskip=13pt
\centerline{\footnotesize\it AECL, Chalk River Laboratories}
\baselineskip=13pt
\centerline{\footnotesize\it Chalk River, Ontario K0J 1J0, Canada}
\centerline{\footnotesize E-mail: hagberg@cu50.crl.aecl.ca}

\vspace*{0.6cm}
\abstracts{Precision studies of $0^{+} \rightarrow 0^{+}$ superallowed
$\beta$ decays provide the necessary data to test stringently the 
CVC hypothesis.  They will also provide a value for the weak vector
coupling constant and the $V_{ud}$ quark mixing element of the
Cabibbo-Kobayashi-Maskawa (CKM) matrix.  
The determination of this element is crucial for the test of
the unitarity of the CKM matrix, and the search for
physics beyond the Standard Model.  This paper reviews the current
status of $0^{+} \rightarrow 0^{+}$ $\beta$-decay data and their 
implications, as well as some relevant data from the $\beta$ decay
of $^{19}$Ne and the neutron.} 

\normalsize\baselineskip=15pt
\setcounter{footnote}{0}
\renewcommand{\thefootnote}{\alph{footnote}}

\section{Introduction}\label{sec:intro}  
Studies of $0^{+} \rightarrow 0^{+}$ superallowed $\beta$ decays are
compelling because of their simplicity. 
The axial-vector decay strength 
is zero for such decays, so 
the measured $ft$ values are directly related to the weak vector 
coupling constant through
the following equation:

\be
ft = \frac{K}{\GV^{\prime 2} \langle \MF \rangle^2} ,
\label{eq:ft}
\ee

\noindent where $K$ is a known constant, $\GV^{\prime}$ is the 
effective vector coupling constant and $\langle \MF \rangle$ is the Fermi
matrix element between analogue states.  Eq.~(\ref{eq:ft}) is only
accurate at the few-percent level since it omits calculated
correction terms.  Radiative corrections, $\delta_{R}$, modify the
decay rate by about 1.5\% and charge-dependent corrections,
$\delta_{C}$, modify the ``pure" Fermi matrix element by about 0.5\%.
Thus, Eq.~(\ref{eq:ft}) is transformed\cite{Ha90} into the
equation:

\be
\F t = ft (1 + \delta_{R}) (1 - \delta_{C}) =
\frac{K}{\GV^{\prime 2} \langle \MF \rangle^2} .
\label{eq:Ft}
\ee

Accurate experimental data on $Q_{EC}$-values, half-lives and branching
ratios combined with the two correction terms then permit precise
tests of the Conserved Vector Current hypothesis, via the constancy
of $\F t$ values irrespective of the $0^{+} \rightarrow 0^{+}$ decay
studied.

These data also yield a value for $\GV^{\prime}$, which in combination 
with the weak vector coupling constant for the purely leptonic muon
decay,
provide a value for $V_{ud}$, the up-down quark mixing element
of the CKM matrix.   Together with the smaller elements, $V_{us}$ and 
$V_{ub}$, this matrix element provides a stringent test of the 
unitarity of the CKM matrix.  Any violation of unitarity would
signal the need for physics beyond the Standard Model, such as
extra $Z$ bosons or the existence of right-handed currents.

\section{The experiments}\label{sec:expts}  
\begin{figure}[t]
\centerline{
\epsfxsize=5.0in
\epsfbox{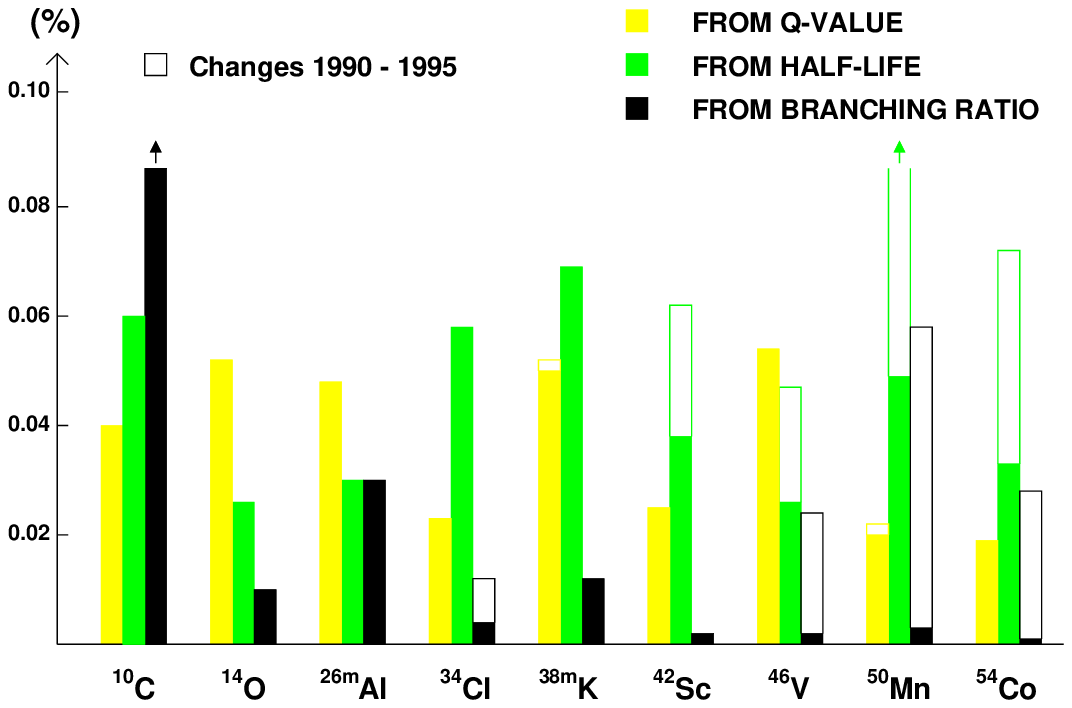}
}
\fcaption{Contributions to the $\F t$-value uncertainty from the
uncertainties in the $Q_{EC}$-values, half-lives and branching
ratios of the nine precisely known $0^{+} \rightarrow 0^{+}$
superallowed beta emitters.  The arrows indicate values that
exceed the scale of the plot.}
\label{fig:fig1}
\end{figure}

The required experimental data on $0^{+} \rightarrow 0^{+}$ beta decays
fall into three categories: {\it (i)} $Q_{EC}$ values, {\it (ii)} 
half-lives and, {\it (iii)} branching ratios.  In order to be useful,
they need to be determined to an accuracy of about 40 parts per million
(ppm) for $Q_{EC}$ and 200 ppm for the other two, a challenge that
requires ingenuity and rigorous procedures.  At present, nine superallowed
beta emitters meet this requirement.  Fig.~\ref{fig:fig1} shows the 
degree to which the necessary experimental data is known in these
nine cases.  Specific examples of precise $Q_{EC}$-value, half-life
and branching-ratio measurements, 
with their associated problems and techniques, are
given below.

\subsection{$Q_{EC}$ values} \label{subsec:qec}
Most precision $Q_{EC}$-value measurements employ $(p,n)$ or
$(^{3}He,t)$ reactions, the inverse beta-decay process, which provide
a simple and direct relation to the beta-decay energy.  Such reaction 
$Q$ values can be determined for the nine well-known superallowed
beta emitters since they all have stable daughter nuclides that will
form the target material in the reaction studies.  In three cases,
$^{26m}$Al, $^{34}$Cl and $^{42}$Sc, 
the nuclide with one less proton than the
$\beta$-decay parent is stable (or long lived) as well.  
Measurements of both the
$(p,\gamma )$ and $(n,\gamma )$ reaction $Q$-values with targets
comprised of these stable nuclides also provide a direct relation to the
$\beta$-decay $Q_{EC}$ value.

In all these measurements the main difficulty lies in calibrating the
projectile and/or ejectile energies.  The Auckland group has
frequently used the $(p,n)$ reaction and eliminated the need for an
ejectile calibration by measuring the sensitive and rapid onset of
neutron emission just at the reaction threshold\cite{Ba92}.  They
calibrate their proton energy by passing the beam through a magnetic
spectrometer, with narrow entrance and exit slits, before it
impinges on the target.  The spectrometer is calibrated before and after
runs with beams of surface ionized K or Cs from an ion source that
is inserted before the spectrometer at a point traversed
by the proton beam.
The source extraction voltage is adjusted until
the alkali beam follows the same path through the spectrometer as
the proton beam and then the energy of the latter 
can easily be deduced from 
the applied extraction voltage.  The spectrometer is NMR stabilized
and never changed between or during runs and calibrations.  The
small final adjustment of the proton beam energy to map out the
$(p,n)$ reaction threshold is done by applying a bias on the target.
The threshold measurements and the calibrations thus all revert to
precise readout of voltages.

At Chalk River the complications of precise, absolute $Q_{EC}$-value
measurements have been avoided by measuring instead the differences
in $Q_{EC}$ values between superallowed beta emitters\cite{Ko87}.
The target material is a mixture of two beta-decay daughter nuclides,
the $(^{3}He,t)$ reaction is used and the outgoing tritons are 
analysed by a Q3D spectrometer. Triton groups originating from both 
types of target nuclei are therefore present in the spectrometer
focal plane.  As in the Auckland measurements the target can be
biased, in this case by up to 150 kV. When a bias of $X$ volts is
applied to the target the incoming, doubly-charged $^{3}$He
projectiles are retarded by $2X$ eV whereas the outgoing, 
singly-charged
tritons are reaccelerated by $X$ eV.  The net effect of the target bias
is a shift of the triton group by $X$ eV along the focal plane.
In these types of $Q_{EC}$-value difference measurements the target
bias is adjusted until the shifted position of a triton peak
from one target component exactly coincides with the unshifted 
position of a triton group from the other component.  When matched,
the trajectories of both triton groups through the spectrometer
are identical and a detailed knowledge of the spectrometer optics
is not required.  The $Q_{EC}$-value difference determination
is then reduced to measuring the target bias. If the two
selected, matched triton groups do not correspond to the final product 
nuclei being in their ground states, 
then a knowledge of the excitation energies
of the reaction products is also required.  
The $Q_{EC}$-value difference measurements
nicely complement the absolute $Q_{EC}$-value measurements and
together they result in a precise $Q_{EC}$-value grid with more than 
one piece of information for each superallowed $\beta$ emitter, 
a situation
that is especially valuable when probing data for possible
systematic biases.

\subsection{Half-lives} \label{subsec:hls}
Measurements of half-lives appear deceptively simple but when high
precision is required they are fraught with possible biases.
Precise half-life measurements of short-lived ($\sim$1 s) nuclides
pose interesting and unique difficulties\cite{Ha92}.  Many samples
need to be counted, with high initial rates, in order to obtain
adequate statistics.  The purity of the radioactive samples must be 
ensured, so they need to be prepared with on-line isotope separators.
This is a major technical challenge because of the low 
volatility and short
half-life of many of the superallowed beta emitters.  The detector
must be capable of high-rate counting and have a high efficiency.
It must also have a very low-energy threshold so as to minimize its
sensitivity to possible, minor gain changes and pile up.  
At high rates the necessary
correction for deadtime becomes the most worrisome aspect of the
counting electronics.  At Chalk River\cite{Ha92} the samples are
prepared with an isotope separator equipped with a high temperature
He-jet ion source. The detector used is a $4\pi$ gas counter with
a 92\% efficiency for positrons.  An instantly retriggerable gate
generator is used to provide a well-defined, non-extendable pulse 
width, which introduces a dominant known deadtime at least five times
longer than any preceeding electronics deadtime.

An accurate result is not guaranteed even if the problems with sample
preparation, detector and electronics have all been addressed because
of the possible bias introduced by the analysis procedure, a source
of potential problems that is often overlooked.  The procedure
employed is based on Poisson statistics, but the counting conditions
do not strictly satisfy the Poisson criteria in that the sample size
is relatively small, often less than $10^4$ atoms, and is counted with
high efficiency until only background is visible.  Furthermore,
the variance for each data point is not the Poisson variance because
of the perturbations introduced by dead time, 
nor is it easily calculable even though the
dead-time losses themselves are properly accounted for.  Numerous
samples are counted in a typical experiment.  Normally, one might 
expect to obtain the final result by averaging the individual
half-lives obtained from a decay analysis of each sample or by
adding together the data from all samples and then performing
a decay analysis, but neither method is strictly correct..

The analysis procedure employed by the Chalk River group uses
modified Poisson variances and is based on a simultaneous fit of up 
to 500 decay curves with a single common half-life, but with individual
intensity parameters for each decay curve.  Because an exact treatment 
is not possible, we have evaluated the bias introduced by our analysis 
simplifications by using hypothetical data generated to simulate closely
the experimental counting conditions.  
The analysis of the hypothetical data should return the same
half-life from which they were generated.
Our analysis has been proven
to be correct at the 0.01\% level.  Our tests with the event generator
have also shown that different analysis procedures, based on
reasonable assumptions, may bias the outcome by more
than 0.1\%, well outside the accuracy required for a $0^{+} \rightarrow
0^{+}$ superallowed beta emitter half-life.

\subsection{Branching ratios} \label{subsec:brs}
The last experimental quantity required is the precise magnitude
of the superallowed branch for each $0^{+} \rightarrow 0^{+}$ emitter.
For eight of the nine well-known cases (excluding $^{10}$C) the
superallowed branch is the dominant one by far and other branches,
if known, are well below the 1\% level.  The non-superallowed
branches seen so far have either been allowed Gamow-Teller transitions 
to $1^{+}$ states in the daughter or been non-analogue transitions to
excited $0^{+}$ states in the daughter.  The latter transitions, 
although very weak, are of special interest because their magnitudes 
are related to the size of one of the necessary calculated 
charge-dependent corrections.  Studies of such non-analogue transitions
are the only way so far to test the predictive power of those
calculations.

For the eight well known $0^{+} \rightarrow 0^{+}$ cases where the
superallowed decay branch is dominating, a precise measurement of
its intensity is achieved by a sensitive, but less precise,
measurement of the weak, competing branches.  The difficulty in
observing these branches stems from the intense background generated
by the prolific and energetic positrons from the superallowed
branch.  At Chalk River a sensitive counting technique has been
developed where events in an HPGe detector, used to observe the
$\gamma$ rays resulting from non-superallowed beta transitions
to excited states in the daughter, are tagged by the direction of
coincident positrons seen in plastic scintillators.  The majority
of unwanted events in the HPGe detector originate from positrons heading
towards the detector since they may interact with it directly or
through bremsstrahlung and annihilation-in-flight processes.  Such
events are removed by a condition that HPGe events must be coincident
with positrons heading away from that detector.  This condition leads to a
dramatic reduction of the background produced by positrons from the
dominant, superallowed ground-state branch, which is not
accompanied by $\gamma$ rays, whereas events from the excited-state
branches, which are accompanied by subsequent $\gamma$ rays, are
still efficiently recorded.  The decays of six superallowed beta
emitters have been investigated with this technique, the results
for four of them have been published\cite{Ha94} so far.  Gamow-Teller
transitions were observed in three cases and non-analogue transitions
in two cases.  Very stringent upper limits for similar branches were
determined for the cases where none was observed.

\section{The theoretical corrections}\label{sec:theo}  
The charged particles involved in a nuclear beta decay interact
electromagnetically with the field from the nucleus as well as with
each other.  These interactions modify the decay when compared to 
a case where only the weak force is involved and they thus need to
be accounted for by theoretical corrections.  For positron decay
the electromagnetic interaction between the proton involved and the
nuclear field, an effect absent in a bare nucleon decay, results in
a charge-dependent correction, $\delta_{C}$, to the Fermi matrix
element.  The similar interaction between the emitted positron
and the nuclear field is already accounted for in the calculation
of the statistical rate function, $f$.  The interactions between the 
charged particles themselves and their associated bremsstrahlung
emissions leads to a series of radiative corrections to the bare
beta-decay process.  It has been found advantageous to group the
radiative corrections into two classes, those that are nuclear-structure
dependent, denoted $\delta_{R}$, and those that
are not, denoted $\Delta_{R}$.

The charge-dependent correction, $\delta_{C}$, arises from the fact
that both Coulomb and charge-dependent nuclear forces act to destroy
the isospin symmetry between the initial and final states in superallowed
beta decay.  The odd proton in the initial state is less bound than the
odd neutron in the final state so their radial wavefunctions differ
and the wavefunction overlap is not perfect.  Furthermore, the degree 
of configuration mixing between the final state and other, excited $0^{+}$
states in the daughter is slightly different from the similar
configuration mixing in the parent, again resulting in an imperfect
overlap.  As was mentioned in Sec.~\ref{subsec:brs} 
the configuration-mixing
predictions have been tested against data on non-analogue $0^{+}
\rightarrow 0^{+}$ transitions. There is good agreement between
the data and the most recent calculations\cite{Ha94,TH95,To95,OB95}.
The radial wavefunction difference correction has been calculated with
Woods-Saxon wavefunctions and the shell model\cite{To95}, the
Hartree-Fock method and the shell model\cite{OB95} and, most
recently, the Hartree-Fock method and the Random Phase 
Approximation\cite{SGS96}.  In general, the three types of 
calculations exhibit similar trends in the predicted values as a 
function of the mass of the emitter, but the absolute values of $\delta_C$
from ref.\cite{To95} differ on average from that of ref.\cite{OB95}
by 0.07\%.

The nuclear-structure dependent radiative correction, $\delta_{R}$,
depends on the energy released in the beta decay and consists of a
series of terms of order $Z^{m} \alpha^{n}$ (with $m < n$) where
$Z$ is the proton number of the daughter nucleus and $\alpha$ is
the fine-structure constant.  The first three terms of this converging
series have been calculated\cite{TH95}. To them is added a
nuclear-structure dependent, order $\alpha$, axial-vector 
term, denoted by $(\alpha / \pi ) C_{NS}$ in ref.\cite{TH95}, 
to form the total
correction, $\delta_{R}$.

The nuclear-structure independent radiative correction, $\Delta_{R}$,
is dominated by its leading logarithm, $(2 \alpha /\pi )  {\rm ln}
(\mZ / m_p)$, where $m_p$ and $\mZ$ are the masses
of the proton and $Z$-boson.
It also incorporates an axial-vector term\cite{TH95},
$(\alpha /2 \pi ) [{\rm ln}(m_p/\mA ) + 2 C_{{\rm Born}}]$,
whose principal uncertainity is the value assigned to the
low-energy cut-off mass, $\mA$.  We adopt\cite{To95} a range
$ \mA /2 < \mA < 2 \mA $ with the central value given by the
$A_1$-resonance mass, $\mA = 1260$ MeV.  The resulting
nucleus-independent radiative correction, $\Delta_{R}$,
then becomes 
$(2.40 \pm 0.08)\%$.

\begin{table} [t]
\protect
\tcaption{Experimental results ($Q_{EC}$, $t_{1/2}$ and branching
ratio, $R$) and calculated corrections ($\delta_C$ and $\delta_R$)
for $0^{+} \rightarrow 0^{+}$ transitions.}
\label{tab:tabl1} 
\footnotesize
\vspace{0.4cm}
\begin{center}
\begin{tabular}{cccccccc}
\hline \\[-3mm]
 & $Q_{EC}$ & $t_{1/2}$ & $R$ & $ft$ & $\delta_C$ &
$\delta_R$ & $\F t$ \\
 & (keV) & (ms) & (\%) & (s) & (\%) & (\%) & (s) \\
\hline \\[-3mm]
$^{10}$C & 1907.77(9) & 19290(12) & 1.4638(22) & 3040.1(51) &
0.16(3) & 1.30(4) & 3074.4(54) \\
$^{14}$O & 2830.51(22) & 70603(18) & 99.336(10) & 3038.1(18) &
0.22(3) & 1.26(5) & 3069.7(26) \\
$^{26m}$Al & 4232.42(35) & 6344.9(19) & $\geq$ 99.97  
& 3035.8(17) &
0.31(3) & 1.45(2) & 3070.0(21) \\
$^{34}$Cl & 5491.71(22) & 1525.76(88) & $\geq$ 99.988  
& 3048.4(19) &
0.61(3) & 1.33(3) & 3070.1(24) \\
$^{38m}$K & 6043.76(56) & 923.95(64) & $\geq$ 99.998  
& 3047.9(26) &
0.62(3) & 1.33(4) & 3069.4(31) \\
$^{42}$Sc & 6425.58(28) & 680.72(26) & 99.9941(14)  
& 3045.1(14) &
0.41(3) & 1.47(5) & 3077.3(24) \\
$^{46}$V & 7050.63(69) & 422.51(11) & 99.9848(13) & 3044.6(18) &
0.41(3) & 1.40(6) & 3074.4(27) \\
$^{50}$Mn & 7632.39(28) & 283.25(14) & 99.942(3) & 3043.7(16) &
0.41(3) & 1.40(7) & 3073.8(27) \\
$^{54}$Co & 8242.56(28) & 193.270(63) & 99.9955(6)  
& 3045.8(11) &
0.52(3) & 1.39(7) & 3072.2(27) \\
\hline
\end{tabular}
\end{center}
\end{table}

\section{Results}\label{sec:resu}  
\begin{figure}[t]
\centerline{
\epsfxsize=4.0in
\epsfbox{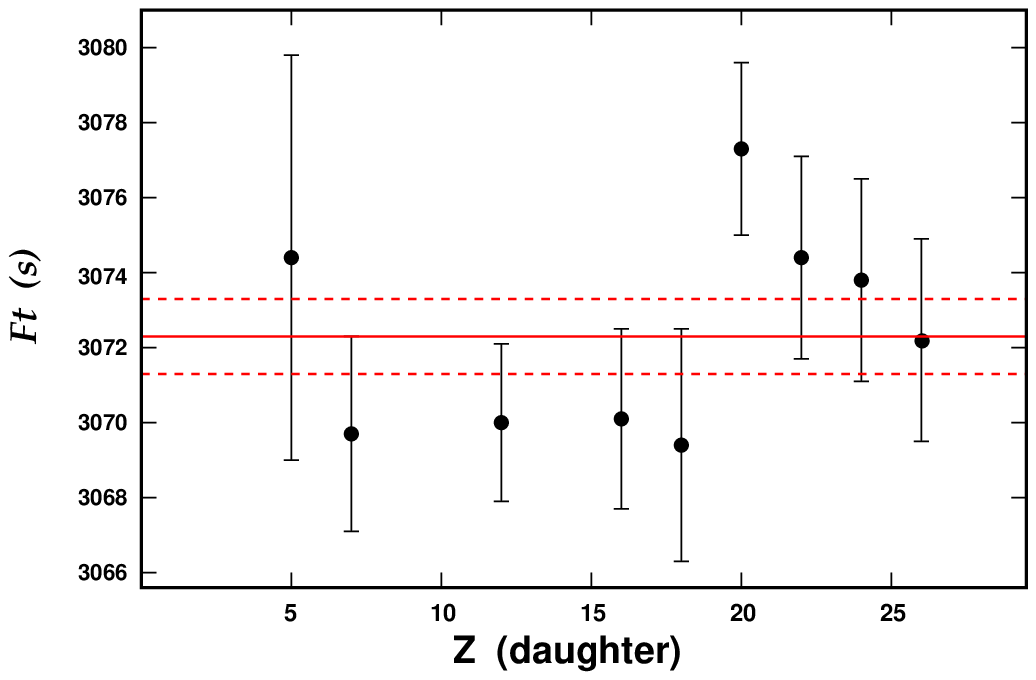}
}
\fcaption{$\F t$ values for the nine well-known cases and the
best least-squares one-parameter fit.}
\label{fig:fig2}
\end{figure}

The measured data on $Q_{EC}$-values, half-lives and branching ratios
for the nine precisely known $0^{+} \rightarrow 0^{+}$ emitters
as well as the calculated charge-dependent and radiative corrections
are given in Table~\ref{tab:tabl1}.  The deduced $\F t$ values for
the nine cases are also shown in Fig.~\ref{fig:fig2}.

It is evident that the nine separate cases are in good agreement,
as is expected from CVC.  The average $\F t$ value is $3072.3 \pm 1.0$~s,
with a reduced chi-square of 1.20.  The constancy of the $\F t$ values
from the nine individual cases establishes that the CVC hypothesis,
as tested through nuclear beta decay, is accurate at the $4 \times 
10^{-4}$ level.

The weak vector coupling constant $\GV^{\prime} = \GV (1 + \Delta_{R}
)^{1/2} = (K/2 \F t)^{1/2}$, deduced from superallowed decay is
$\GV^{\prime}/(\hbar c)^3 = (1.1496 \pm 0.0004) \times 10^{-5}$
GeV$^{-2}$, where the error on the average $\F t$ value has been
doubled\cite{To95} to include an estimate of the systematic
uncertainties in the calculated correction, $\delta_{C}$.  
The $V_{ud}$ quark mixing element of the CKM matrix
is defined as $V_{ud} = \GV / G_{\mu}$, where $G_{\mu}$ is the
coupling constant deduced from the purely leptonic muon decay.
We arrive at $V_{ud} = 0.9740 \pm 0.0005$.  With values of the
other two elements of the first row of the CKM matrix taken
from ref.\cite{PDG96} the unitarity test produces the following
result

\be
\mids V_{ud} \mids^2 +
\mids V_{us} \mids^2 +
\mids V_{ub} \mids^2 = 0.9972 \pm 0.0013 .
\label{eq:utnuc}
\ee

\begin{figure}[t]
\centerline{
\epsfxsize=5.0in         
\epsfbox{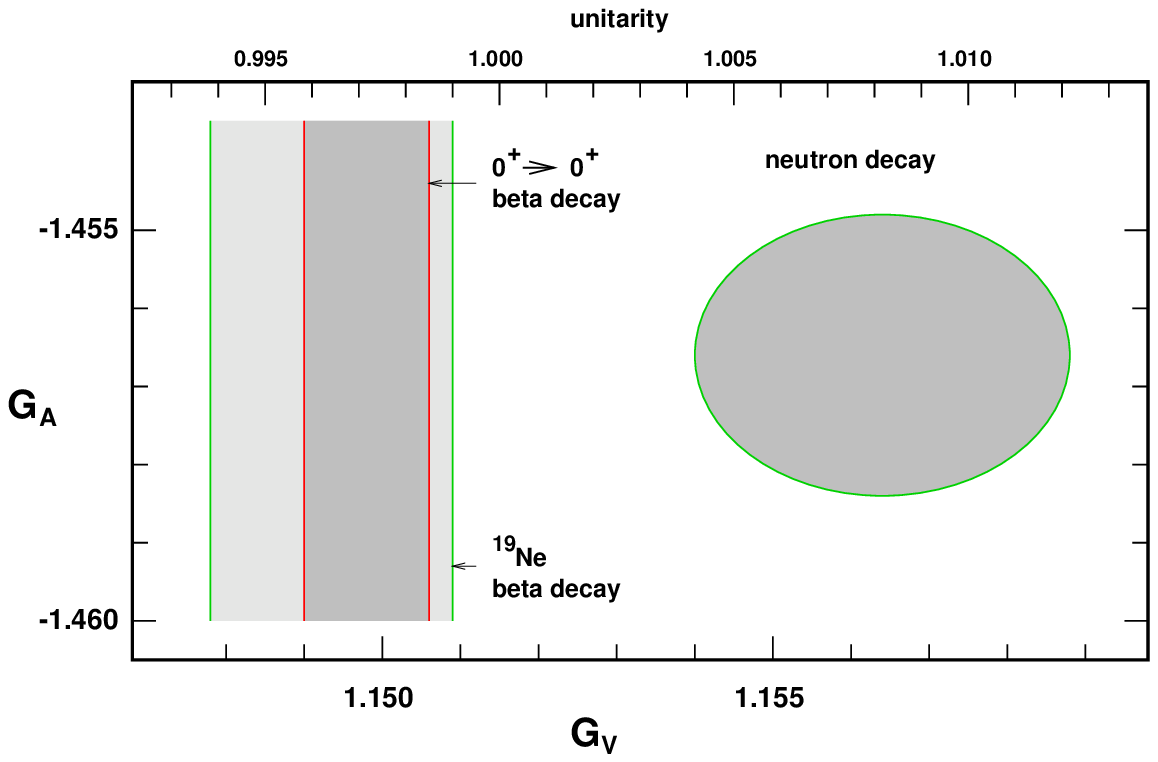}
}
\fcaption{Allowed regions of the vector coupling constant and
axial-vector coupling constant from nuclear superallowed beta
decays and neutron decay.  The scale at the upper part of the figure 
translates the $\GV$ scale into the corresponding unitarity sum
of the first row of the CKM matrix.}
\label{fig:fig3}
\end{figure}

\noindent The discrepancy with unity, shown in Fig.~\ref{fig:fig3},
is more than two standard deviations.

Precision measurements of non $0^{+} \rightarrow 0^{+}$ superallowed
beta decays can also yield $\F t$ and $V_{ud}$ values.  The decays
of $^{19}$Ne and the neutron have been studied extensively but because
their superallowed decay branches contain Gamow-Teller components,
beta-asymmetry measurements are required to separate them from the
Fermi components.  Consequently the precision achieved so far in these 
two cases is less than that achieved in the $0^{+} \rightarrow 0^{+}$
decay measurements. The first row unitarity test with a $V_{ud}$
value based on the most current $\beta$-decay asymmetry measurement
of $^{19}$Ne\cite{Jo96} is also shown in Fig.~\ref{fig:fig3}.
It is in good  agreement with the $0^{+} \rightarrow 0^{+}$
data, which further supports the CVC hypothesis but still leaves a
unitarity problem.  
The $^{19}$Ne data also yields a result for $\GA \langle \MGT \rangle
$, which can be used to deduce a value for the axial-vector coupling
constant, $\GA$.  However, unlike the case of the neutron,
the Gamow-Teller matrix element, $\langle \MGT \rangle$, is not
explicitly given by theory but needs to be calculated, for example,
with the shell model.  Consequently a very high precision is not
attainable for $\GA$ from $^{19}$Ne decay studies and only the
$\GV$ value is shown in Fig.~\ref{fig:fig3}.
The results for $\GV$ and $\GA$ from the neutron decay
studies\cite{TH95} are also shown in Fig.~\ref{fig:fig3}.  With
$V_{ud}$ based on the neutron studies the unitarity test also fails,
but now the sum of the matrix elements is too large.

The current status is thus far from ideal.  All three types of data,
nuclear superallowed beta decay, $0^{+} \rightarrow 0^{+}$ and non
$0^{+} \rightarrow 0^{+}$, and the decay of the neutron, result in a 
failure to meet the unitarity condition.  Only two of the three
types of data agree among themselves.  However, it is worth
pointing out that the different types of data have their own
particular strengths and weaknesses.  The strength of the
superallowed decay studies is the large number of cases, which dilute 
the effect of any possible, erroneous measurement, and their consistency.
The weakness is the number, magnitude and complexity of the necessary,
calculated corrections.  It is unlikely to expect a large change in the
$\F t$ value deduced from superallowed beta emitters from further
experimental or theoretical work.

In contrast, the strength of the neutron decay studies is the
simplicity of the calculated corrections.  The weakness is that
it is a single case with fewer measurements and, consequently a
greater susceptibility to one possible erroneous measurement.
(In fact, the two most recent $\beta$-asymmetry measurements
disagree.)   The neutron case thus appears to have greater potential
but it is also the one most likely to see its $\F t$ value
move substantially from its present location.

\end{document}